# A magnonic directional coupler for integrated magnonic half-adders

**Q. Wang[1,2*], M. Kewenig[2], M. Schneider[2], R. Verba[3], F. Kohl[2], B. Heinz[2,4], M. Geilen[2], M. Mohseni[2], B. Lägel[5], F. Ciubotaru[6], C. Adelmann[6], C. Dubs[7], S. D. Cotofana[8], O. V. Dobrovolskiy[1], T. Brächer[2], P. Pirro[2], and A. V. Chumak[1,2*]**

[1]*Faculty of Physics, University of Vienna, Boltzmanngasse 5, A-1090 Vienna, Austria*

[2]*Fachbereich Physik and Landesforschungszentrum OPTIMAS, Technische Universität Kaiserslautern, D-67663 Kaiserslautern, Germany*

[3]*Institute of Magnetism, Kyiv 03142, Ukraine*

[4]*Graduate School Materials Science in Mainz, Staudingerweg 9, 55128 Mainz, Germany*

[5]*Nano Structuring Center, Technische Universität Kaiserslautern, D-67663 Kaiserslautern, Germany*

[6]*Imec, B-3001 Leuven, Belgium*

[7]*INNOVENT e.V., Technologieentwicklung, Prüssingstraße 27B, 07745 Jena, Germany*

[8]*Department of Quantum and Computer Engineering, Delft University of Technology, Delft, 2600 The Netherlands*

*Magnons, the quanta of spin waves, could be used to encode information in beyond-Moore computing applications, and magnonic device components, including logic gates, transistors, and units for non-Boolean computing, have already been developed. Magnonic directional couplers, which can function as circuit building blocks, have also been explored, but have been impractical because of their millimetre dimensions and multi-mode spectra. Here, we report a magnonic directional coupler based on yttrium iron garnet single-mode waveguides of 350 nm width. We use the amplitude of a spin-wave to encode information and to guide it to one of the two outputs of the coupler depending on the signal magnitude, frequency, and the applied magnetic field. Using micromagnetic simulations, we also propose an integrated magnonic half-adder that consists of two directional couplers and processes all information within the magnon domain with aJ energy consumption.*

[*] Corresponding authors: qi.wang@univie.ac.at, andrii.chumak@univie.ac.at

Spin waves (and their quanta magnons) in magnetic structures could potentially be used as data carriers in future low-energy computing devices [1-5]. Spin waves can transfer information with low losses [1-4,6,7] and can be used to implement logic functionality based on a wide range of nonlinear spin-wave phenomena [8-10]. The phase of a coherent spin wave provides an additional degree of freedom (beyond amplitude) in data processing, decreasing the footprint of logic units [11-13], and magnonic structures can be scaled down to the nanometre regime [14,15] and using spin waves with nanometre wavelengths [16-18]. Nanoscale single-mode magnonic waveguides can also overcome the issue of parasitic magnon scatterings into higher modes [19]. Furthermore, reducing the dimensions of magnonic structures to the atomic scale could potentially shift the frequency of the spin waves from the GHz to the THz range [20,21].

Several magnon-based data processing devices have already been demonstrated: spin-wave logic gates [11,22-25], majority gates [26,27], magnon transistors and valves [8,28], spin-wave multiplexers [29,30], and unconventional and neuromorphic computing elements [31-33]. However, an integrated all-magnonic circuit, which is suitable for the cascading of multiple magnonic units, has not yet been developed. Nanoscale spin-wave directional couplers with reconfigurable functionality can constitute the core of integrated all-magnonic circuits [34]. But while spin-wave directional couplers have been explored experimentally [35], their millimetre dimensions and multi-mode spectrum limit their practical implementation.

In this Article, we report a magnonic directional coupler based on yttrium iron garnet (YIG) which allows for single-mode operations due to its sub-micrometer waveguide widths. Using space-resolved micro-focused Brillouin light scattering (μBLS) spectroscopy [36], we investigate its functionality as a building block for integrated magnonic circuits. In the linear regime, the directional coupler exhibits the functionality of a microwave filter for the processing of analogue and digital information, a power splitter for fan-out logic gates, and a frequency divider or signal multiplexer. In the nonlinear regime, the outputs of the directional coupler can be controlled by varying the spin-wave amplitude, which can be useful for logic gates. We also combine linear and nonlinear directional couplers numerically to construct a half-adder - a prototype of a magnonic integrated circuit. Numerical benchmarking of the proposed half adder (based on 30 nm technology) against a 7 nm complementary metal–oxide–semiconductor (CMOS) half-adder shows that the proposed device has a 10-fold lower energy consumption and a comparable device footprint.

**Magnonic directional coupler structure**

Our sub-micrometre directional coupler (Fig. 1a) is fabricated from an 85 nm thick YIG film [6,7] (see Methods) and consists of two spin-wave waveguides with a width of 350 nm. Near the point of spin-wave excitation, the waveguides are physically separated by a narrow gap of 320 nm. To transfer spin waves out of the coupled waveguides into an “isolated” conduit, the waveguides bend at an angle of 12-degrees until a gap of 1.32 μm is reached. A U-shaped antenna is placed on top of the first YIG waveguide to excite spin waves, and at a distance of 2 μm to the second waveguide to avoid spin-wave excitation in both waveguides (see Extended Data Fig. 1 and Supplementary Note 1). When a field of 56 mT is applied along the waveguides, spin-wave frequencies ranging from 3.4 GHz to 3.63 GHz are well excited by the U-shaped antenna in the first waveguide (see Extended Data Fig. 2 and Supplementary Note 2). Only the first width mode can be excited in this frequency range (single-mode nano-waveguide) as shown in the dispersion curve (Fig. 1b). To detect the spin-wave intensity in the directional coupler, space-resolved μBLS spectroscopy is used (see Methods) [36].

**Linear functionality**

As a first step, we measure spin-wave intensity in five points along each output waveguide as marked by blue and red crosses in Fig. 1a. Figure 1c shows the spin-wave intensities for the two output waveguides averaged over these points as a function of the excitation frequency. As it can be seen, the two spectra show quite different features: In the first waveguide, the maximum spin-wave intensity was observed at 3.58 GHz. In contrast, the maximum intensity in the second waveguide was found around 3.465 GHz and only very weak spin-wave intensities are detected above 3.575 GHz. To understand the nature of this frequency separation, the dispersion relations of the first two spin-wave width modes for coupled waveguides are shown in Fig. 1b (see Methods). The color-coding represents the results of micromagnetic simulations, whereas the dashed lines are calculated using analytical theory (see Methods) [15,19]. The dispersion curve of the first width mode splits into antisymmetric (as) and symmetric (s) modes due to the dipolar interaction between the waveguides. This results in an oscillation of the spin-wave energy between the coupled waveguides [19,35]. Thus, once the spin-wave energy is injected into only one of the waveguides, it will be transferred entirely to the other one after the propagation of a certain distance, which is known as the coupling length $L$. It is defined by the wavenumbers of the spin-wave modes $k_{as}$ and $k_s$: $L = \pi / \Delta k = \pi / |k_{as} - k_s|$ and depends strongly on the spin-wave frequency and other parameters [19]. Since the length of the coupled waveguides is fixed, the ratio of this length to the coupling length $L$ defines, in which of the two output waveguides of the directional coupler the spin wave is guided. Figure 1d shows the frequency dependence of the normalized output spin-wave intensities for both output waveguides. The experimental data are well fitted by the developed analytical model (see Methods), indicating the high robustness of the proposed directional coupler design. The measured maximal transfer of the spin-wave energy takes place at a spin-wave frequency of around 3.48 GHz and is equal to 93.8%, which is only slightly below the theoretical value of 100%. This difference is likely due to imperfections of the fabricated structure and might be decreased by the further improvement of the nanostructuring process [37].

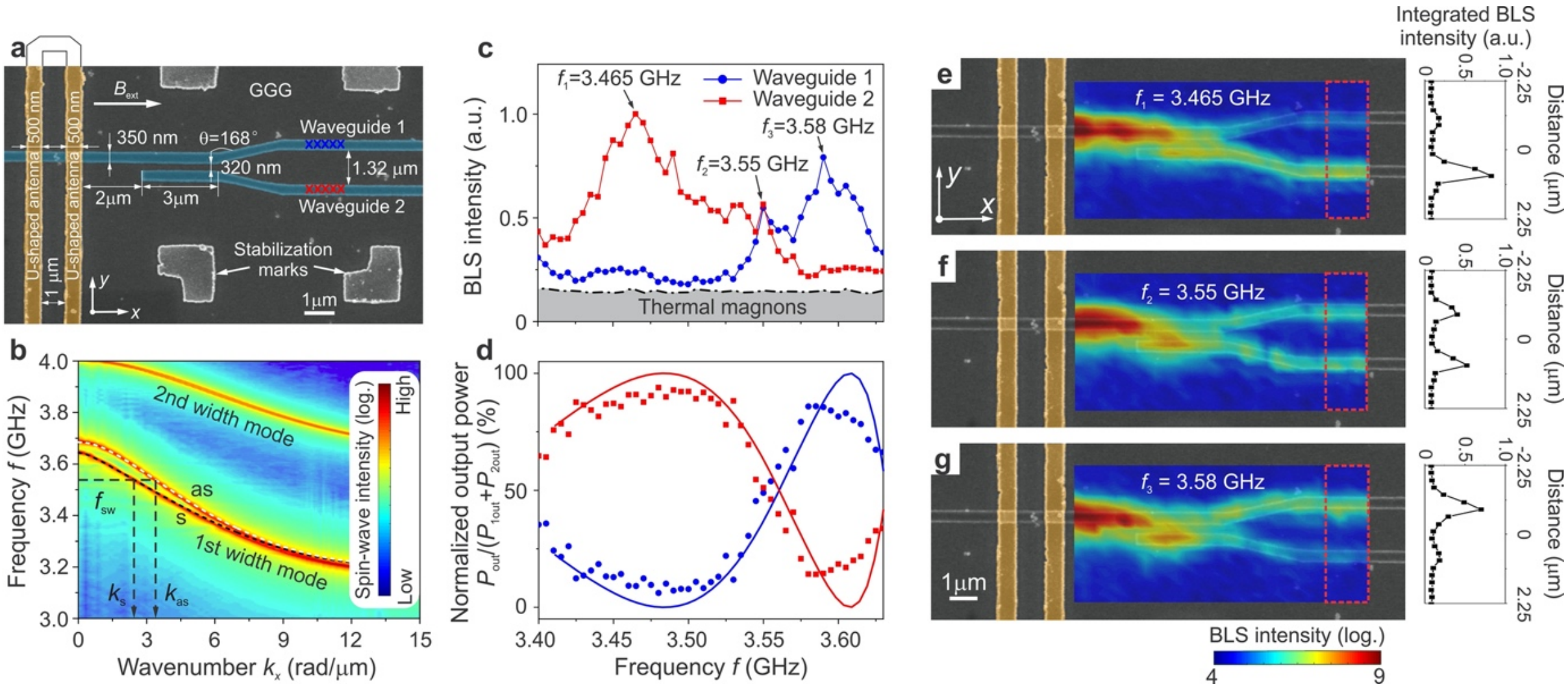

***Fig. 1. Sample geometry and working principle of the directional coupler in the linear regime. a***. *Scanning electron microscopy (SEM) image of the directional coupler (shaded in blue) with the U-shaped antenna. External magnetic field* $B_{ext}$ *= 56 mT is applied along the YIG conduits (x-axis) to saturate the directional coupler in a backward volume geometry [34] and radio-frequency (RF) current with power* $P_{mw}$ *= 0 dBm is applied to the antenna in order to excite spin waves.* ***b***. *Spin-wave dispersion relation of the first two width modes obtained using micromagnetic simulation (color-coded) and analytic theory (dashed lines). YIG waveguides of 350 nm width and 320 nm gap in between are considered.* ***c***. *Averaged spin-wave spectra measured by μBLS spectroscopy on the first (blue circles) and on the second (red squares) output waveguide. The arrows indicate the frequencies, which are chosen for the demonstration of different functionalities of the directional coupler in Fig. 1****e-g***. ***d***. *The frequency dependence of the normalized output powers* $P_{out}/(P_{1out}+P_{2out})$ *with subtracted thermal background for both waveguides. Circles and squares represent experimental results and solid lines are theoretical calculations of the normalized output spin-wave intensity at the first (blue) and the second (red) output waveguide.* ***e-g***. *Two-dimensional BLS maps (the laser spot was scanned over an area of 9.4×4.5 μm*$^2$ *by 30×20 points) of the BLS intensity for* ***e***. $f_1$ *= 3.465 GHz,* ***f***. $f_2$ *= 3.55 GHz, and* ***g***. $f_3$ *= 3.58 GHz. The right panels show the spin-wave intensity integrated over the red dashed rectangular regions at the end of the directional coupler.*

It should be emphasized that complex magnonic circuits are only possible using single-mode waveguides. In these waveguides, width modes are well-separated in energy to prevent elastic inter-mode scatterings [38]. The parasitic scatterings would introduce an energy loss for the signal-carrying mode and create complex interference patterns due to the simultaneous presence of waves with different wave vectors. It is especially critical for the concepts based on directional couplers since different wave vectors possess also different coupling lengths. The waveguides used in our studies are single-mode due to their nanoscopic sizes which ensures the modes separation (see Fig. 1b). An additional advantage of the nanoscopic waveguides is spin-wave propagation in longitudinally self-magnetized waveguides allowing for the efficient 2D guiding of information [19].

Two-dimensional BLS spectroscopy scans of the spin-wave intensity are shown in Fig. 1(e-g) to directly demonstrate the frequency-dependent functionality of the directional coupler. Figure 1e shows the case, where most of the spin-wave energy is transferred to the second waveguide at a spin-wave frequency of 3.465 GHz. Thus, this planar two-dimensional directional coupler can be used to efficiently connect two magnonic conduits without the need for complex and costly three-dimensional bridges, which are used in modern electronic circuits. Figure 1g shows an entirely different spin-wave path in the directional coupler. The increase in the spin-wave frequency up to $f_3$ = 3.58 GHz results in a decrease of the coupling length $L$ by roughly a factor of two. As a result, the spin wave transfers all its energy from the first waveguide to the second one and back. Thus, 86% of the total output spin-wave energy is guided back into the first output waveguide of the directional coupler. This demonstrates the potential use of the directional coupler as a frequency division demultiplexer: If different frequencies are applied to the same input of the directional coupler, they will be transferred to the different outputs of the device. Finally, Figure 1f demonstrates that the directional coupler can also be used as a 50/50 power splitter, in which half of the spin-wave energy is transferred to the second waveguide and half of it remains in the first one. Such a splitter can also be used as a fan-out logic gate if an amplifier [39-41] will be installed at the outputs of the device to compensate the split in energy. Furthermore, for a fixed frequency, the output signal of the directional coupler can be switched from one output to the other by changing the external field in a small range of $\Delta B_{\text{ext}}$ = 4.7 mT (see Extended Data Fig. 3 and Supplementary Note 3). Thus, magnetic fields from switchable nanosized magnets [42] could be used to realize a non-volatile ns-fast reconfigurability of the directional coupler.

**Nonlinear switching functionality**

The processing of data, in general, requires the utilization of elements with nonlinear characteristics that are, e.g., provided by a semiconductor transistor in CMOS. As mentioned above, the key benefits of spin waves for the data processing are their pronounced natural nonlinearity that allows for an all-magnon control of one magnonic unit by another. In our studies, the phenomenon of a nonlinear shift of the dispersion relation [9,10] is used in contrast to the multi-magnon scattering exploited in the realization of a magnon transistor [8]. In the relatively weak nonlinear regime, where the dipolar coupling between the waveguides is larger than the nonlinear frequency shift of the spin waves, the nonlinear operation of the directional coupler can be described simply by taken into account the nonlinear frequency shift of the symmetric and antisymmetric collective modes. The shift is the same for both modes can be well approximated by the nonlinear frequency shift of waves in isolated waveguides [43]: $f_{\text{s,as}}^{(nl)}(k_x, a_k) = f_{\text{s,as}}^{(0)}(k_x) + T_k |a_k|^2$, where $f_{\text{s,as}}^{(0)}(k_x)$ are the dispersion relations of the symmetric and antisymmetric modes of the coupled waveguides in the linear region [19], $a_k$ is the canonical spin-wave amplitude and $T_k$ is the nonlinear shift coefficient (see Methods). For the backward volume geometry ($M_s // k_x$) used here, the nonlinear shift coefficient is negative [9,43] and, thus, the spin-wave dispersion curves shift down with an increase in the spin-wave amplitude defined by the applied RF power. The calculated spin-wave dispersions are shown in Fig. 2a for small and large applied microwave powers. As it can be seen, for a fixed spin-wave frequency of 3.52 GHz, the coupling length $L$ decreases from $\pi/\Delta k_x^{\mathrm{lin}}$ to $\pi/\Delta k_x^{\mathrm{nonlin}}$ with an increase in input power, resulting in changed device characteristics.

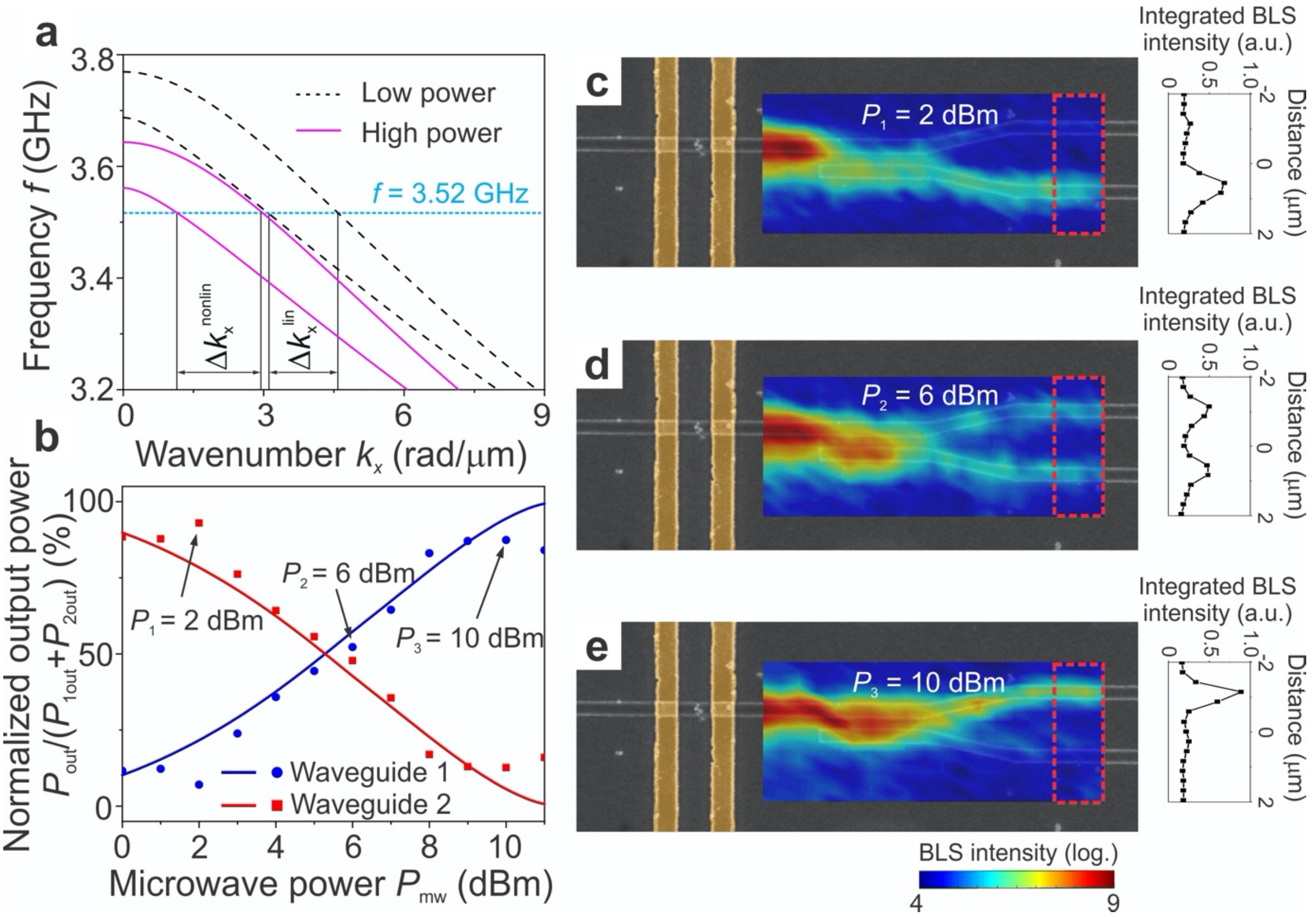

*Fig. 2* ***Nonlinear functionality of the directional coupler. a****. The dispersion relations of symmetric and antisymmetric spin-wave modes in the coupled waveguides for small (black dashed lines) and large (magenta solid lines) powers. The increase in spin-wave amplitude results in the downshift of the dispersion curves.* ***b****. Averaged output spin-wave intensity as a function of the microwave power $P_{mw}$. (dots – experimental results, lines – theoretical fit)* ***c-e****: Two-dimensional BLS maps of the spin-wave intensity for a frequency of f = 3.52 GHz and different input powers* ***c.*** *$P_1$ = 2 dBm,* ***d.*** *$P_2$ = 6 dBm, and* ***e.*** *$P_3$ = 10 dBm. The right panels show the spin-wave intensity integrated over the regions indicated by red dashed rectangular.*

To study the nonlinear switching functionality of the presented directional coupler, the microwave power $P_{\mathrm{mw}}$ was varied in the range from 0 dBm to 11 dBm. Figure 2b clearly shows that the respective output spin-wave intensity strongly depends on the input microwave power due to the discussed nonlinear effects. Figure 2c shows that for a relatively low input power 2 dBm, the output spin-wave energy is transferred to the second waveguide. This regime can be considered as a linear one. For the increased power of 6 dBm, the spin-wave dispersion shift implies that half of the output spin-wave energy is transferred back to the first waveguide and thus, the directional coupler, thus, works as a 50/50 splitter. A further increase of the input power up to 10 dBm results in a further dispersion shift, a decrease of the coupling length $L$ and in a transfer of the spin-wave energy back to the first waveguide as it can be seen in Fig. 2e.

## Design of all-magnon half-adder

According to obtained experimental results, we propose an integrated magnonic circuit on the example of a half-adder consisting of two directional couplers and investigate its functionality using the means of micromagnetic simulations. The simulations allow us to check the working principle of this design in the size comparable to the CMOS device and also to perform the benchmarking. In the simulations, we have chosen the minimal width of the waveguides to be 100 nm (see Fig. 3 for the sizes of the structure), which can be reliably fabricated using modern patterning techniques [14,15,37,44] (see Methods).

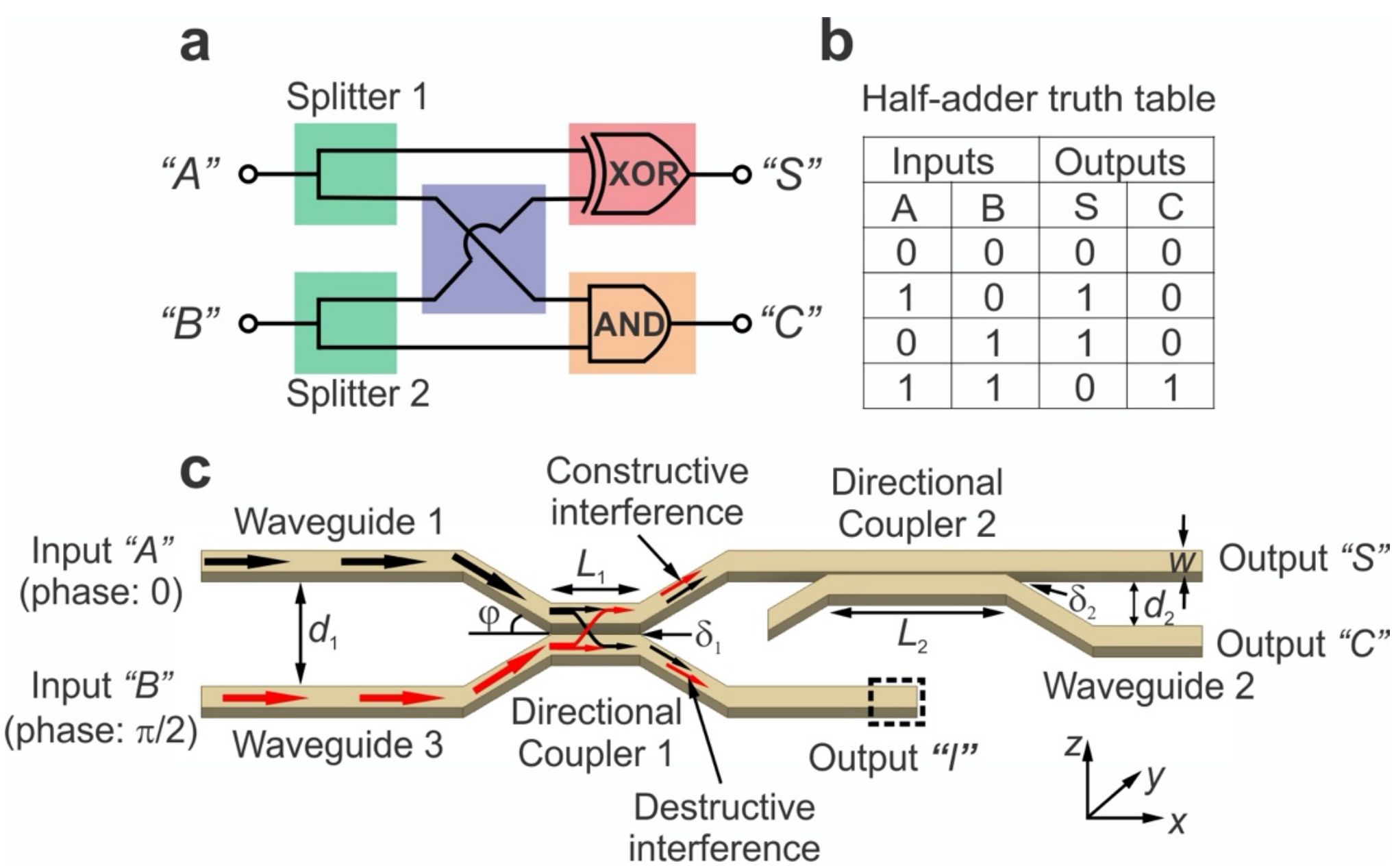

| Inputs | | Outputs | |
|---|---|---|---|
| A | B | S | C |
| 0 | 0 | 0 | 0 |
| 1 | 0 | 1 | 0 |
| 0 | 1 | 1 | 0 |
| 1 | 1 | 0 | 1 |

***Fig. 3 The operational principle of the magnonic half-adder**. **a** Sketch of half-adder in electronics. Building blocks are highlighted by different colors. **b** Half-adder truth table. **c** Schematic view of the magnonic half-adder. In this work we consider the following parameters: The widths of the YIG waveguides are w = 100 nm, thicknesses are h = 30 nm, edge to edge distances between different waveguides are $d_1$ = 450 nm, $d_2$ = 210 nm, the angles between the waveguides are $\varphi$ = 20°, the gaps between the coupled waveguides are $\delta_1$ = 50 nm, $\delta_2$ = 10 nm, and the lengths of the coupled waveguides are $L_1$ = 370 nm and $L_2$ = 3 μm. The arrows show the magnons flow path from inputs to logic gates.*

A general schematic layout of a half-adder in electronics is shown in Fig. 3a. It combines an XOR logic gate and an AND logic gate using three-dimensional bridge constructions. It adds two single binary digital Inputs "*A*" and "*B*" and has two Outputs, sum ("*S*") and carry ("*C*"). The truth table of a half-adder is shown in Fig. 3b. The sketch of the proposed magnonic half-adder is shown in Fig. 3c. Directional Coupler 1 in the magnonic half-adder acts as a power splitter for each of the two inputs and, at the same time, replaces the three-dimensional bridge required for sending the signals from Input "A" to the gate AND and from Input "B" to the gate XOR (compare Fig. 3a). The spin-wave flow paths in the magnonic half-adder are shown by the black and red arrows in Fig. 3c: Spin waves from both inputs are split into two identical spin waves of half intensity by the Directional Coupler 1. One pair of the waves is directed to the Directional Coupler 2 via the Waveguide 1 and the other pair is guided into the idle Output "*I*" via the Waveguide 3. In the present simulation, the Output "*I*" just features a high damping region at the end (shown in the figure with a dashed rectangle) and it does not contribute to the half-adding function. However, it acts as an XOR logic gate and, with the use of another directional coupler, can perform the same half-adder operation (see Extended Data Fig. 4 and Supplementary Note 4). Thus, the modified half-adder can be considered as a combination of a half-adder with a fan-out logic gate, which doubles each output of the device. The Directional Coupler 2 performs the actual half-adder logic operation and its operational principle is described in the next section.

**Modelling of the nonlinear functionality**

The nonlinear functionality of the directional coupler shown above qualitatively takes place for any spin-wave directional coupler. Nevertheless, as it is shown below, the realization of the logic operation requires the full switch of the spin-wave path by the change in the spin-wave intensity exactly four times. To achieve this value, the modifications of the directional coupler discussed below are required.

The Directional Coupler 2 consists of coupled straight parallel waveguide with 3 μm coupled length as shown in Fig. 4a. The split dispersion relations in the linear regime in the coupled waveguides are shown in Fig. 4b by the blue lines. To obtain the linear dispersion, small spin-wave amplitudes are excited by a microwave field of $h_{rf} = 2$ mT. The output power in the first waveguide normalized by the total power $P_{1out}/(P_{1out}+P_{2out})$ can be expressed using the characteristic coupling length $L_{C2}$: $\frac{P_{1out}}{P_{1out}+P_{2out}} = \cos^2\left(\pi L_2/\left(2L_{C2}\right)\right)$, where $L_2 = 3$ μm is the length of the coupled waveguide in the Directional Coupler 2. Figure 4c shows the normalized output power in the first waveguide as a function of the spin-wave frequency $f$ in the frequency range from 2.28 GHz to 2.65 GHz. The result of numerical simulations in the linear regime is shown with the blue symbols and the analytic calculation with the solid blue lines. One can clearly see that the output power $P_{1out}$ strongly depends on the spin-wave frequency as it has been shown experimentally in Fig. 2. This is due to the strong dependence of the coupling length $L_{C2}$ on the spin-wave wavenumber [19,35,45]. The coupling length, consequently, defines the energy distribution between the output waveguides for a given length of the coupled waveguides. The small mismatch between simulations and theory in the region below 2.3 GHz is mainly caused by the damping, which is not taken into account in the theory, and by the large sensitivity of the coupling coefficient to the dispersion of the antisymmetric mode, which is practically flat in this region.

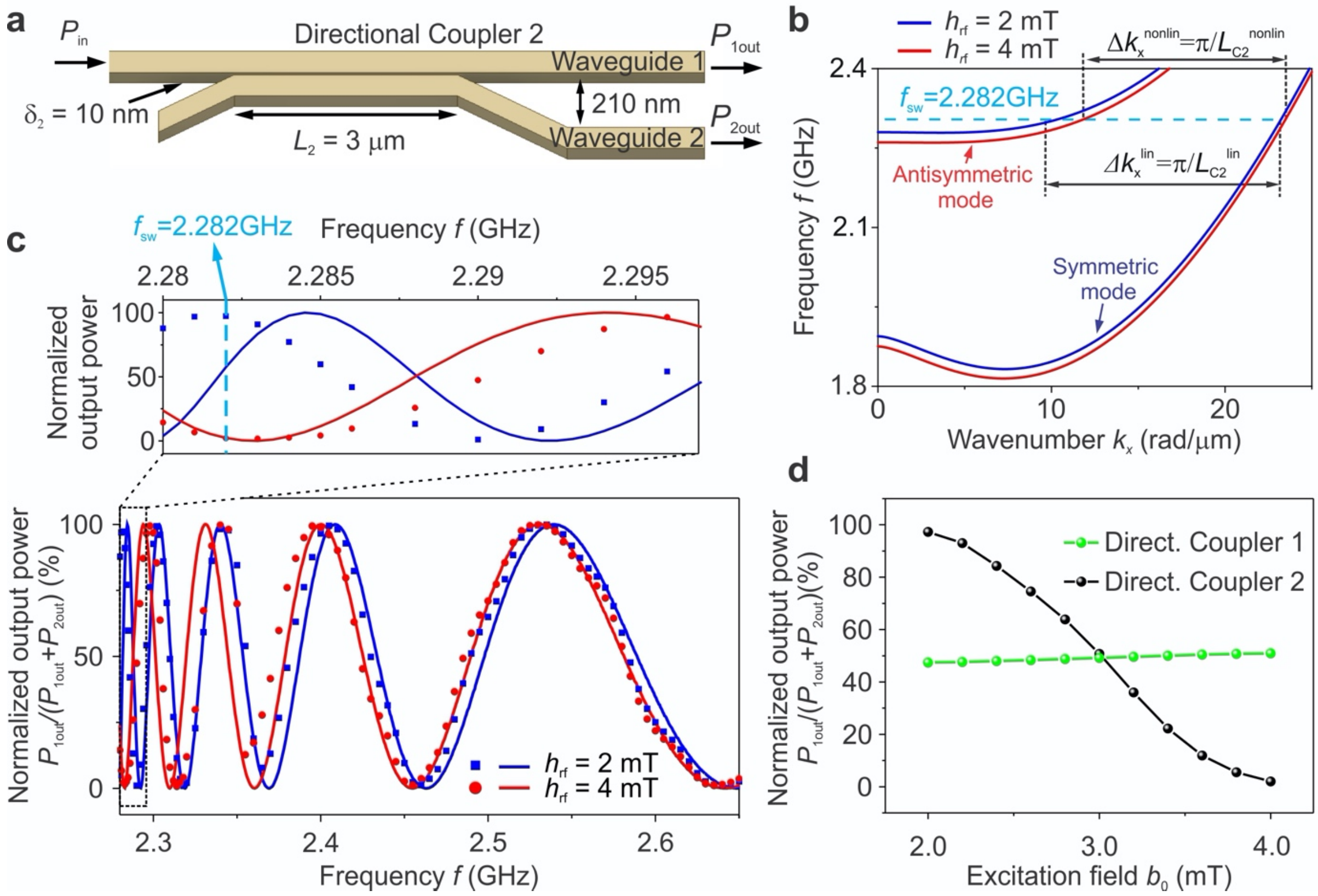

***Fig. 4 Modelling and characteristics of the Directional Coupler 2.*** ***a*** *Schematic of the directional coupler 2.* ***b*** *Analytically-calculated dispersion curves for the coupled waveguides for small (blue lines) and large (red lines) excitation fields* $h_{rf}$*. The change in the coupling length* $L_{C2}$ *is clearly visible that is associated with the increase in spin-wave amplitude.* ***c*** *Normalized output power in the first waveguide* $P_{1out}/(P_{1out}+P_{2out})$ *as a function of frequencies for different excitation field* $h_{rf}$ *(symbols - simulations, lines – analytic theory). A zoom of the region marked with the dashed rectangle is shown in the top panel.* ***d*** *Simulated normalized output power* $P_{1out}$ *as a function of the excitation field* $b_0$ *for a fixed frequency of* $f$ *= 2.282 GHz for Directional Coupler 1 and 2.*

When the input spin-wave power increases, spin-wave dispersion shifts down – see red lines in Fig. 4b. Thus, a fixed spin-wave frequency of 2.282 GHz becomes correspondent to different spin-wave wavenumbers what changes the parameter $\Delta k_x = \pi/L_{C2}$ from $\Delta k_x^{\mathrm{lin}}$ to $\Delta k_x^{\mathrm{nonlin}}$ with an increase in the excitation field from $b_0$ = 2 mT to 4 mT. Consequently, the coupling length $L_{C2}$ of the directional coupler also changes. Using the Taylor expansion of the frequency dependence of the coupling length, the power dependence of the output of Directional Coupler 2 can be found:

$$\frac{P_{1\mathrm{out}}}{P_{1\mathrm{out}}+P_{2\mathrm{out}}} = \cos^2\left(\frac{\pi L_2}{2L_{C2}^{\mathrm{lin}}} - \frac{L_2}{L_{C2}^{\mathrm{lin}}}\frac{\pi}{2L_{C2}^{\mathrm{lin}}}\frac{\partial L_{C2}}{\partial f} T_k \left|a_k\right|^2\right) \tag{1}$$

The power-independent term is proportional to the ratio of the directional coupler length to the coupling length in the linear regime $L_2/L_{C2}^{\mathrm{lin}}$. The output power $P_{1\mathrm{out}}$ periodically changes with a change in the coupling length and is maximal for the cases $L_2/L_{C2}^{\mathrm{lin}} = 0, 2, 4, \dots$ (see Fig. 4c). Simultaneously, as it is seen from Eq. (1), the sensitivity to the nonlinear effect increases with the increase in the ratio $L_2/L_{C2}^{\mathrm{lin}}$.

Therefore, the longer the directional coupler is and the more coupling lengths it spans, the higher is the nonlinear phase accumulation. This is the reason why the Directional Coupler 2 in our half-adder design is long and features a strong coupling provided by the small gap between the waveguides of only 10 nm. It has a length of $L_2 = 14L_{C2}^{\text{lin}}$ and it is very sensitive to the increase in the spin-wave amplitude passing through it. As a result, a complete energy transfer from Output 1 to Output 2 is observed in the micromagnetic simulations if the spin-wave intensity is increased by a factor of four ($L_2 = 13L_{C2}^{\text{nonlin}}$) – see the black line in Fig. 4d. The normalized output spin-wave power in the first waveguide decreases from 97.3% at $b_0 = 2$ mT to 2.0% at $b_0 = 4$ mT. Due to this nonlinear switching effect, the Directional Coupler 2 performs a combined AND and XOR logic function, as will be described in the following. At the same time, the first Directional Coupler 1 in the half-adder design should remain in the linear regime and its coupling length should be independent on the spin-wave power. This is achieved by its smaller length of 370 nm as well as via an increased spacing between the waveguides of 50 nm. As a result, Directional Coupler 1 spans only half of the coupling length $L_1 = 0.5L_{C1}^{\text{lin}}$ independent on the excitation field – see green symbols in Fig. 4d. The directional coupler studied experimentally above was designed for the linear functionality. Nevertheless, the increase of the spin-wave power from 0 dBm to 10 dBm, which is more that 4 times increase in the spin-wave intensity required by the half-adder design, also results in the nonlinear switch – see Fig. 2.

**Operational principle of the magnonic half-adder**

The operational principle of the half-adder is shown in Fig. 5. Binary data is coded into the spin-wave amplitude, namely, in the ideal case, a spin wave of a given amplitude (e.g., $M_z/M_s = 0.057$) corresponds to the logic state "1" while zero spin-wave amplitude corresponds to "0". In the following, we normalize all output spin-wave intensity to the input spin-wave intensity. In the more realistic cases considered below, we utilize an approach from CMOS: a normalized spin-wave intensity below 1/3 is considered to be logic "0" and above 2/3, it is logic "1".

The operational principle of the half-adder is as follows. Let us first consider the case of logic inputs "*A*" = "1" and "*B*" = "0" – see Fig. 5a. In this case, the spin wave injected into the Input "*A*" is split into two equal parts by the Directional Coupler 1. One of them is directly guided to the Directional Coupler 2 by the upper conduit. The spin-wave intensity is chosen in such a way that Directional Coupler 2 remains in the linear regime ($L_{C2}^{\text{lin}} \approx 214$ nm $\approx 14/L_2$) and after initial oscillations, the spin wave is guided into the Output "*S"* as shown in Fig. 5a. Only about 1.9% of the spin-wave energy goes into the Output "*C"*. This corresponds to the logic Outputs "*S"* = "1" and "*C"* = "0". If a spin wave is injected in Input "*B"* only, this corresponds to the logic inputs *A*" = "0" and "*B*" = "1" – see Fig. 5b. The situation in this case is quite similar to the previous one. The situation is different for the input logics states "*A"* = "1" and "*B"* = "1" – see Fig. 5c. It is assumed that the phase of the spin wave injected into the Input "*B*" is permanently shifted by π/2 with respect to the one in the Input "*A*"(in order to compensate the -π/2 phase shift caused by the Directional Coupler 1), which can be easily realized by many means [46]. In this case, constructive interference of both spin waves will take place in Waveguide 1 and destructive interference in Waveguide 3. As a result of this coherent superposition, the entire spin-wave energy from both inputs goes to Directional Coupler 2 resulting in four times larger spin-wave intensity with respect to the single-input cases when only 50 % of the spin-wave energy is guided to this coupler. As was discussed above, this increase in the spin-wave intensity by a factor of four switches the coupler to the nonlinear regime ($L_{C2}^{\mathrm{nonlin}} \approx 230$ nm $\approx 13/L_2$) and the spin wave is guided to the output "*C*". This corresponds to the logic Outputs "*S*" = "0" and "*C*" = "1" (see Fig. 5c) and, thus, the whole truth-table of the half-adder is realized.

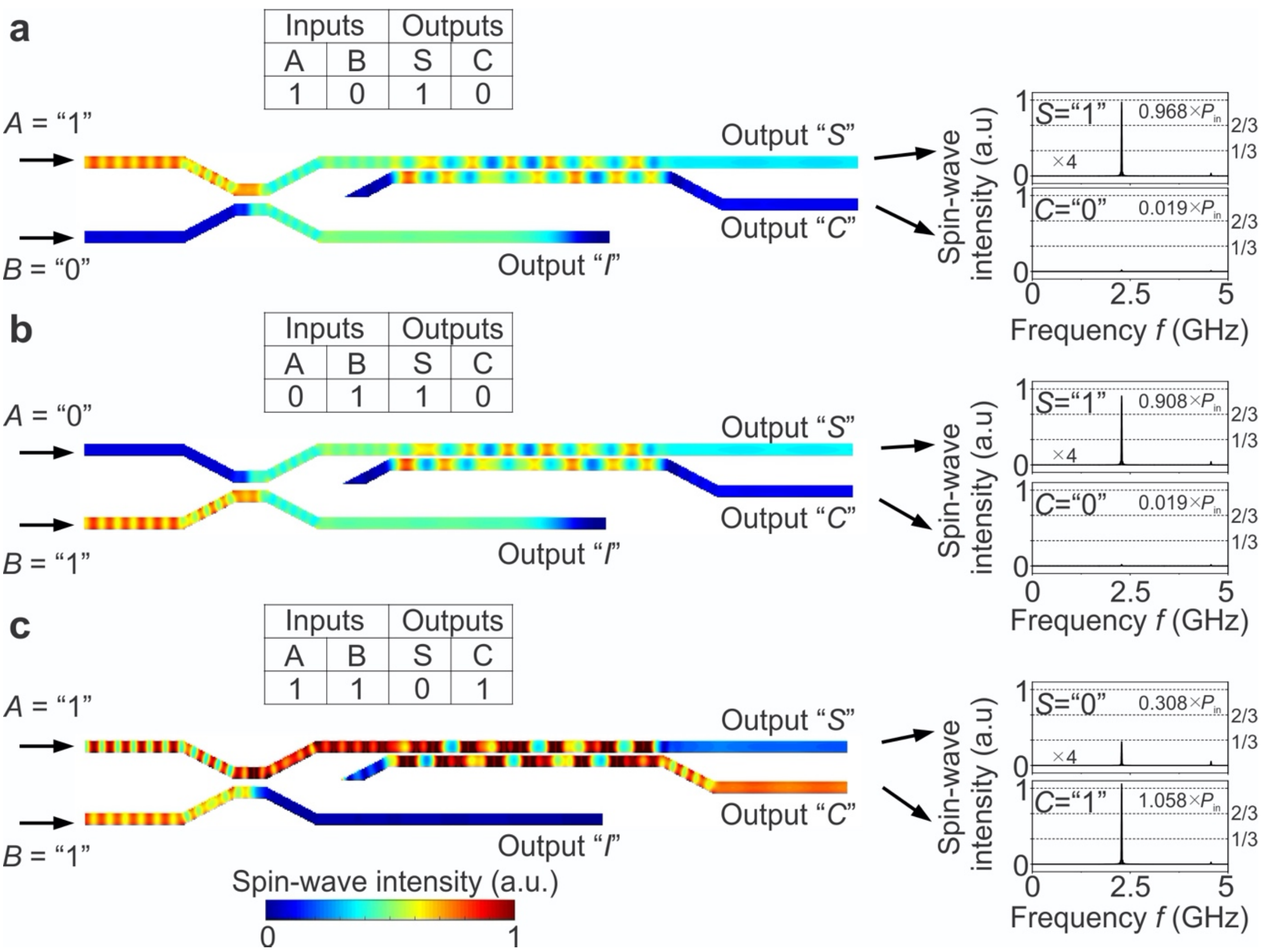

***Fig. 5 Operational principle of the magnonic half-adder.*** *The spin-wave intensity distributions for different inputs combinations:* ***a*** *"A" = "1", "B" = "0",* ***b*** *"A" = "0", "B" = "1" and* ***c*** *"A" = "1", "B" = "1". The truth tables for each input combination are shown on the top of the structure. The normalized output spin-wave intensities in the outputs are shown on the right side. All the outputs "S" are multiplied by a factor of four.*

Please note that the all-magnon circuit concept [8,34] requires that the signal from the output of a magnonics gate can be directly guided into the input of the next one. In order to satisfy this condition, the spin-wave intensity at the Outputs "*S*" still has to be amplified approximately by a factor of four due to the energy splitting in the Directional Coupler 1 and due to parasitic reflections and spin-wave damping in the waveguides. The output signals "*S*" shown in Fig. 5 are artificially multiplied by 4. The most promising realization of such an amplifier is based on the utilization of voltage controlled magnetic anisotropy (VCMA) parametric pumping [43,47,48] (see Extend data Fig. 5 and Supplementary Note 5). In contrast, no amplifier is required for the "Carry" output of the half-adder. In general, the idea presented here and the concept of the half-adder is applicable for any magnetic material. Nevertheless, the requirement that the device length $L_{de}$ is smaller than the spin-wave decay length should be satisfied. This is the case not only for YIG, but also for low damping Heusler compounds [49].

**Benchmarking of the proposed technology**

The summary of the key parameters of two versions of the proposed half-adder is shown in Table 1 (see Methods): The first one is the device that was simulated and discussed above. The second device is an estimation performed for a device with $w$ = 30 nm, $h$ = 10 nm and minimal gap $\delta$ = 10 nm. It has to be mentioned that the second device does not constitute a fundamental limit but is merely an estimation based on the current state of the art of fabrication technology [15,37]. A further improvement in all characteristics is potentially achievable.

**Table 1. Magnonic half-adder benchmarking.**

| Parameters | YIG * (100 nm) | YIG † (30 nm) | CMOS ‡ (7 nm) |
|---|---|---|---|
| Area ($\mu m^2$) | 5.58 | 1.016 | 1.024 |
| Delay time (ns) | 150 | 18 | $6\times10^{-2}$ |
| Total energy consumption without amplification (aJ) | 24.6 | 1.96 | 35.3 |
| Spin-wave frequency (GHz) | 2.282 | 2.29 | - |
| Spin-wave wavelength (nm) | 340 | 510 | - |
| Spin-wave group velocity (m/s) | 25 | 137 | - |

| Type of amplifier | Energy consumption |
|---|---|
| Electric current based parametric pumping [39,41] | $10^5$ (aJ/operation) |
| Voltage controlled magnetic anisotropy parametric pumping [43,47] | 3 (aJ/operation) |

***Footnote:*** **The values in this column are extracted from the micromagnetic simulation of the half-adder investigated in the paper. † The values are estimated characteristics of a device miniaturized down to 30 nm using the Eqs. (8) - (10) (see Methods). ‡ The values are calculated using Cadence Genus by Sorin D. Cotofana for 7 nm CMOS technology.*

According to the table, the area of the simulated 100 nm feature size half-adder is 5.58 $\mu m^2$ (the spaces between the neighbouring logic gates are also included) and is, thus, only a few times larger than a corresponding 7 nm feature size CMOS device. In contrast to a CMOS realization, the magnonic half-adder core part (without amplifier) consists of only three nano-wires made of the material and of only one planar layer. This drastically simplifies its fabrication and decreases its potential costs. The area can be readily decreased down to 1.016 $\mu m^2$ for the second 30 nm-based device, which is comparable to the 7 nm-based CMOS device. In addition, it should be noted that the largest part of the half-adder is given by the Directional Coupler 2, which could potentially be further decreased via the utilization of exchange coupling mechanisms between the waveguides instead of dipolar coupling. To achieve this, the air gap between the coupled waveguides should be filled with another magnetic material.

Operational frequency is an important requirement. In the presented half-adder, the delay time is defined by the whole length of the device with respect to the spin-wave group velocity. In our design, the spin-wave propagation time from input to output is about 150 ns. According to Table 1, the calculation time can be reduced to 18 ns in the second device. This value is larger than the 60 ps delay time obtained for 7 nm CMOS and assumes that magnon logic would be more suitable for slow but low-energy applications. At the same time, one has to note that CMOS does not operate on its maximal speed because of the drastically growing Joule heating (a typical clock rate is 3 GHz, which corresponds to around 0.3 ns delay).

Small energy consumption of computing systems is probably the most crucial requirement taking into account the constantly increasing amount of information that has to be processed. In our simulations, we record the total energy of the device as a function of simulation time. The energy injected into the device per nanosecond is equal to $4.1\times10^{-20}$ J/ns for the input combinations "*A*" = 1, and "*B*" = 1. Please note that only the energy propagating along the positive direction is taken into account. For the 300 ns pulse duration the energy consumption is, thus, 12.3 aJ. For all the operations, the total energy consumption is 24.6 aJ. This is similar to current CMOS values (35.3 aJ), which is calculated using Cadence Genus (see Methods). It should be especially highlighted that the energy consumption of the miniaturized 30 nm-based device is more than one order of magnitude smaller and is around 1.96 aJ. At the same time, we have to underline that this energy consumption is related to the energy within the magnonic domain only and the energy consumption of the amplifier should be added (see Table 1 and Supplementary Note 5). The most promising approach is VCMA parametric pumping that has been reported recently [43,47,48] and allows for the energy consumption of an amplifier ~ 3 aJ per device.

### Conclusions

We fabricated a sub-micrometre spin-wave directional coupler operating in a single-mode regime and studied its functionality in the linear and nonlinear regime using μBLS spectroscopy. Our experimental results are supported by numerical simulations and an analytical theory. By varying the applied microwave frequency or an applied magnetic field, spin-waves can be guided to different coupler outputs, demonstrating the reconfigurability of the device. Our spin-wave directional coupler could therefore be potentially used as a microwave filter for processing analogue and digital information, a power splitter for fan-out logic gates, a frequency divider or signal multiplexer, and a planar interconnecting element for magnonic conduits. Furthermore, the output of the directional coupler can be switched by changing the spin-wave amplitude, demonstrating nonlinear functionality.

We have also proposed and tested numerically an integrated magnonic circuit - a half-adder - based on the fabricated directional coupler. This half-adder consists of two directional couplers: one that functions as a linear power splitter and one that functions as a nonlinear switch (as demonstrated experimentally). The proposed device is all-magnonic - the magnons are controlled by magnons without any conversion to the electric domain - ensuring low energy consumption. The proposed magnonic half-adder consists only of three planar magnetic nanowires with one amplifier and can potentially substitute 14 transistors in electronics circuits. A magnonic half-adder developed with 30 nm technology is predicted to have a footprint comparable to a 7 nm CMOS half-adder and around ten times smaller energy consumption.

## Methods

#### Liquid phase epitaxial film growth and sample fabrication.

A 85 nm thick Yttrium Iron Garnet (YIG) film has been grown on a 1-inch (111) 500 µm thick Gadolinium Gallium Garnet (GGG) substrate by liquid phase epitaxy from PbO-$B_2O_3$ based high-temperature solutions at 860°C using the isothermal dipping method (see e.g. Ref. [50]). Nominally pure $Y_3Fe_5O_{12}$ films with smooth surfaces were obtained on horizontally rotated substrates applying rotation rates of 100 rpm [6,7]. The saturation magnetization of the YIG film is $1.42\times10^5$ A/m and its Gilbert damping $\alpha = 2.1\times10^{-4}$ as it was extracted by ferromagnetic resonance spectroscopy [51].

The directional coupler was fabricated by using electron beam lithography, $Ar^+$ ion beam etching and electron beam evaporation. First a double layer of polymethyl methacrylate (PMMA) was spin coated on the YIG film and the directional coupler structures were created afterwards by using electron beam lithography. To get well shaped waveguides, Titanium and Chromium were deposited by electron beam evaporation as a bilayer hard mask, which defines the shape of the directional coupler structures. These were then etched out of the film by $Ar^+$ ion beam etching. Finally, the U-shaped antenna was defined by using electron beam lithography and a lift-off process. It consists of ~ 230 nm thick gold and 20 nm thick titanium (for adhesion).

#### BLS spectroscopy and spin-wave excitation.

Micro-focused Brillouin Light Scattering (µBLS) spectroscopy is a technique for the measurements of spin-wave intensities with frequency, space, phase, and time resolution [36]. It is based on the inelastic light scattering of the incident laser beam by magnons. In our measurements, a laser beam of 491 nm wavelength and a power of 1.8 mW is focused on the directional coupler with an effective spot diameter of 400 nm using a ×100 microscope objective with a large numerical aperture (NA=0.75). The scattered light was collected and guided into a tandem Fabray-Perot interferometer TFP-1 (JRS Scientific Instruments) for further analysis. To perform the two-dimensional scans, the sample was moved with respect to the laser spot in steps of a few hundred nanometre in each direction using a piezoelectric stage. The stabilization marks were grown on the sample to be able to keep the same relative position of the laser spot during the long measuring cycles.

An external magnetic field $B_{ext}$ = 56 mT is applied along the YIG conduits (*x*-axis) to saturate the directional coupler in a backward volume geometry [34] and a radio-frequency (RF) current with power $P_{mw}$ = 0 dBm (in the linear regime) is applied to the antenna in order to excite spin waves. The spin-wave frequencies ranging from 3.4 GHz to 3.63 GHz are well excited by the U-shaped antenna in the isolated waveguide (See Extended Data Fig. 2 and Supplementary Note 2). Only the first width mode is excited in this frequency range as shown in the dispersion curve (Fig. 1b).

**Calculation of nonlinear frequency shift coefficient.**

The nonlinear shift coefficient $T_k$ in the isolated waveguide can be calculated using the framework of [9] and by assuming a uniform mode profile across the waveguide thickness and width. Accounting for the negligible static demagnetization of a waveguide along its length, $F_0^{xx}=0$, the nonlinear shift coefficient becomes equal to [40]:

$$T_k=\left(\left(\omega_H-A_k\right)+\frac{B_k^2}{2\omega_0^2}\left(\omega_M\left(4\lambda^2k_x^2+F_{2k}^{xx}(0)\right)+3\omega_H\right)\right)\Big/2\pi, \tag{2}$$

where

$$A_k=\omega_H+\frac{\omega_M}{2}\left(2\lambda^2k_x^2+F_k^{yy}(0)+F_k^{zz}(0)\right), \tag{3}$$

$$B_k=\frac{\omega_M}{2}\left(F_k^{yy}(0)-F_k^{zz}(0)\right). \tag{4}$$

The relation between the dynamic magnetization component and the canonical spin-wave amplitude $a_k$ is given by:

$$M_z=M_s a_k\sqrt{2-\left|a_k\right|^2}\left(u_k-v_k\right), \tag{5}$$

with

$$u_k=\sqrt{\frac{A_k+\omega_0}{2\omega_0}}\ \text{and}\ v_k=-\operatorname{sign}\left[B_k\right]\sqrt{\frac{A_k-\omega_0}{2\omega_0}}. \tag{6}$$

**Calculation of the directional coupler characteristics.**

The theory of the directional coupler has been described in our previous paper [19]. However, in Ref. [19] we accounted only for the straight part of the coupler where the distance between waveguides is minimal and constant. For the device reported in this article, this approach is not sufficient, since the gap between the coupled waveguides is quite large (320 nm). In such a case, the region of the bent waveguides could also significantly contribute to the coupling characteristics, since the gap in this region is not much larger than the minimal gap over a considerable distance. To take this bent region into account, we calculated the splitting of the symmetric and antisymmetric spin-wave modes as a function of the gap, $\Delta k=\Delta k\left(d\right)$. Then, the coordinate dependence of the spin-wave power in the waveguides is given by

$$\begin{aligned}P_1(x)&=\cos^2\left[\int_0^x\frac{1}{2}\Delta k\left(d\left(x'\right)\right)dx'\right],\\ P_2(x)&=\sin^2\left[\int_0^x\frac{1}{2}\Delta k\left(d\left(x'\right)\right)dx'\right].\end{aligned} \tag{7}$$

Furthermore, an additional coupling, which is especially pronounced for large spin-wave wavelengths, must be taken into account due the large wavelength studied in this work: The part of the first waveguide located before the second one starts also contributes to the coupling "diagonally". Indeed, the dynamic magnetization of a large spin-wave wavelength varies slowly and, thus, the mentioned part of the first waveguide creates a non-negligible dipolar field at the beginning of the second one. In contrast, for short wavelength spin waves, these additional contributions vanish, because contributions from neighbouring half-wavelength parts almost cancel each other. In this work, we account for it by the introduction of an "additional effective length" of the coupler which, by itself, depends on the spin-wave wavelength. Since the strength of the dipolar fields decays with the distance approximately proportional to $x^{-3}$, the effective length is expected to depend on spin-wave wavenumber as $L_{\text{eff}}=C_1\Big/\left(k+C_2\right)^2$. Here, the second power in spin-wave wavenumber comes from the integration $\int_0^{1/k}\left(x+d_0\right)^{-3}dx$ and the constant $C_2$ reflects the fact that the effective length cannot increase infinitely for an infinitely large spin wave wavelength. By fitting the experimental data, we found $C_1$ = 25 μm$^{-1}$ and $C_2$ = 2 μm$^{-1}$. Using this expression for the effective additional length, both, frequency and field dependencies of the power transmission rates are well described (see Fig. 1d and Fig. 2b).

The variation of the power transmission rate in the coupler with increasing spin-wave power is mainly attributed to the nonlinear frequency shift of the symmetric and antisymmetric spin-wave modes in coupled waveguides, as shown in Fig. 3b. The shift of the dispersion results in a change of spin-wave wavenumbers at given frequency and, consequently, in a change of the coupling between the waveguides. Knowing the frequency dependence of the power transmission rates $P_{1,2}^{(\mathrm{lin})}(\omega)$ in the linear regime, the nonlinear characteristics can be calculated simply as $P_{1,2}^{(\mathrm{nl})}(\omega,a)=P_{1,2}^{(\mathrm{lin})}\left(\omega-T_k|a|^2\right)$, where $a$ is the canonical spin-wave amplitude and $T_k$ is the nonlinear frequency shift ( $T_k/2\pi=-1.8$ GHz in our case ). Since experimental data measured for 0 dBm excitation power also corresponds to a weakly nonlinear regime, for the description of power dependence we use the relation $P_{1,2}^{(\mathrm{nl})}(\omega,a)=P_{1,2}^{(0)}\left(\omega-T_k\left(|a|^2-|a_0|^2\right)\right)$, where $P_{1,2}^{(0)}$ - dependence for 0 dBm (measured and fitted by calculations above) and $a_0$ - spin-wave amplitude at 0 dBm excitation power. The relation of the spin-wave amplitude with the excitation power was obtained by measuring the BLS intensity in the first waveguide before the coupler and fitting one adjusting parameter (ratio of BLS counts to the square of the spin-wave amplitude). We get the following relation $a=0.035\sqrt{1+p/17.4}$ , where $p$ is the excitation power in dBm. The appearance of an almost linear dependence of the spin wave power on $p$, instead of an exponential one, which could be expected, is mediated by the strong variation of the spin-wave group velocity with spin-wave wavenumber, and, consequently, with the excitation power at a given frequency. The described simple model fits the experimental data well for the applied powers below 10 dBm (see Fig. 3a). For the higher powers, higher-order nonlinear effects should be taken into account additionally [52].

**Micromagnetic simulations**.

(Dispersion curve presented in Fig. 1b) The micromagnetic simulations were performed by the GPU-accelerated simulation program Mumax3 to calculate the space- and time-dependent magnetization dynamics in the investigated structures using a finite-difference discretization [53]. The following material parameters are used: The saturation magnetization $M_s$ = 1.33×10$^5$ A/m is 94% comparing to the value of the plain film [6,7] due to the $Ar^+$ ion beam etching and Gilbert damping is α = 2×10$^{-4}$. A standard exchange constant of YIG A = 3.5 pJ/m was assumed. There were three steps involved in the calculation of the spin-wave dispersion curve [54]: (i) The external field was applied along the waveguide, and the magnetization was relaxed to a stationary state (ground state). (ii) A sinc field pulse $b_y$=$b_0$sinc(2π$f_c$t), with an oscillation field $b_0$ = 1 mT and a cut-off frequency $f_c$ = 10 GHz, was used to excite a wide range of spin waves. (iii) The spin-wave dispersion relations were obtained by performing the two-dimensional fast Fourier transformation (FFT) of the time- and space-dependent data.

(Magnonic half-adder) The simulated structure of the magnonic half-adder is shown in Fig. 1c. The parameters of nanometer thick YIG are obtained from experiment and are as follows [6,7]: saturation magnetization $M_s$ = 1.4×10$^5$ A/m, exchange constant $A$ = 3.5 pJ/m, and Gilbert damping $\alpha$ = 2×10$^{-4}$. The damping at the ends of the simulated structure and the high damping absorber is set to exponentially increase to 0.5 to prevent spin-wave reflection [55]. The high damping region could be realized in the experiment by putting another magnetic material or metal on top of YIG to enhance the damping or it can just correspond to waves guided into further parts of the magnonic network. No external bias field is applied. The static magnetization orients itself parallel to the waveguides spontaneously due to the strong shape anisotropy in the nanoscale waveguides. The mesh was set to 10×10×30 nm$^3$. To excite propagating spin waves, a sinusoidal magnetic field $b$ = $b_0$sin(2π$ft$) is applied over an area of 100 nm in length, with a varying oscillation amplitude $b_0$ and the microwave frequency $f$. $M_z(x,y,t)$ of each cell was collected over a period of 300 ns, which is long enough to reach the steady state. The fluctuation $m_z(x,y,t)$ were calculated for all cells via $m_z(x,y,t) = M_z(x,y,t) - M_z(x,y,0)$, where $M_z(x,y,0)$ corresponds to the ground state. The spin-wave spectra of the output signals are calculated by performing a fast Fourier transformation from 250 ns to 300 ns, which corresponds to the steady state. We would like to mention that all these simulations were performed for the defect-free waveguides and without taking into account temperature. The influences of the edge roughness, trapezoidal cross-sections of the waveguides, and temperature can be ignored due to their smallness as it has been shown in our previous studies [15,19].

**Energy consumption**

For the estimation of energy consumption in the magnonic system (neglecting transducers), the minimal energy consumption can be express as (see Supplementary Note 6):

$$E = \frac{20\pi}{3}\frac{M_s}{\gamma}\frac{v_{gr} f S}{T_k} \tag{8}$$

where $v_{gr} = 2\pi\frac{\partial f}{\partial k}$ is the spin-wave group velocity, $S$ is the cross-section of the waveguide. As one can see, the energy consumption is independent of the characteristics of spin-wave couplers and spin-wave amplitude. Note that the nonlinear frequency shift $T_k$ is of the order of the spin-wave frequency $f$ ($T_k \propto f$), especially in the exchange-dominated region. The conclusion has arrived that the feasible way to reduce the energy consumption is the decrease of waveguide cross-section $S$. Another alternative is searching for specific points or mechanisms with anomalously high nonlinearity. It should be noted that the relation of the Eq. (8) is universal and takes place in other realizations of magnonic half-adders, which are based on the nonlinear shift. For the other designs, the only change is the pre-factor 20π/3.

**Scalability and delay time**

The width of the device can be estimated by:

$$w_{de} = 2w + 4\times 5h \tag{9}$$

where $w$ is the width of waveguide and $h$ is the thickness of waveguide. This equation accounts for the minimal distance between all waveguides and neighboring devices at $5h$ to make dipolar interaction relatively weak. The gaps between different logic gates are taken into account in this width.

The length of the device is given by

$$L_{de} = \left(N + 0.5\right)L_C + 4\frac{5h}{\sin\varphi} \tag{10}$$

where $\varphi$ is the angle of the bent waveguide, $L_C$ is the coupling length, $N = L_2/L_C$ is the ratio between the coupled length of Directional Coupler 2 and the coupling length. The minimal $N$ can be estimated from the condition that Directional Coupler 1, working at half the coupling length, does not significantly change its characteristics at power, which is sufficient to switch the Directional Coupler 2. Simple calculations yield that the change of Directional Coupler 1 transmission is given by $\cos^2\left(\pi\left(N-1\right)/\left(4N\right)\right)$, while in the linear regime transmission rate is equal to 1/2. This gives the restriction $N_{min} = 6$. The area of the magnonic half-adder is $Area = w_{de}L_{de}$. The processing delay is $\tau_d = L_{de}/v_{gr}$.

**Calculation of energy consumption of 7 CMOS half-adder**

We considered a 7nm HA standard cell afferent to the typical processor corner (room temperature, 0.7V power supply) and evaluated its power consumption using Cadence Genus. To this end, we set an inverter standard cell as driver and a capacitance of 2.5fF as output load, and assume for the nets a 50% probability of logic "1" and a toggle rate of 0.02 per ns. Simulation results indicate a total power consumption of 587.994nW out of which the dynamic component (divided into nets power and internal power, which account for 87.7% and 13.3% of the dynamic power, respectively) dominates the less than 1nW leakage component.

**Data availability**

The data that support the plots within this paper and other findings of this study are available from the corresponding author upon reasonable request.

**References**

1 Kruglyak, V. V., Demokritov, S. O. & Grundler, D. Magnonics. *J. Phys. D Appl. Phys.* **43**, 264001 (2010).

2 Lenk, B. *et al.* The building blocks of magnonics. *Phys. Rep.* **507**, 107-136 (2011).

3 Krawczyk M. & Grundler D., Review and prospects of magnonic crystals and devices with reprogrammable band structure. *J. Phys.-Cond. Matt.* **26**, 123202 (2014).

4 Chumak, A. V. *et al.* Magnon spintronics. *Nat. Phys.* **11**, 453-461 (2015).

5 Dieny, B *et al.* Opportunities and challenges for spintronics in the microelectronic industry. *arXiv: 1908.10584* (2019).

6 Dubs, C. *et al.* Sub-micrometer yttrium iron garnet LPE films with low ferromagnetic resonance losses. *J. Phys. D Appl. Phys.* **50**, 204005 (2017).

7 Dubs, C. *et al.* Low damping and microstructural perfection of sub-40nm-thin yttrium iron garnet films grown by liquid phase epitaxy. *Phys. Rev. Materials* **4**, 024426 (2020).

8 Chumak, A. V., Serga, A. A., & Hillebrands, B. Magnon transistor for all-magnon data processing. *Nat. Commun.* **5**, 4700 (2014).

9 Krivosik, P. & Patton, C. E. Hamiltonian formulation of nonlinear spin-wave dynamics: Theory and applications. *Phys. Rev. B* **82**, 184428 (2010).

10 Sadovnikov, A. V. *et al.* Nonlinear spin wave coupling in adjacent magnonic crystals. *Appl. Phys. Lett.* **109**, 042407 (2016).

11 Khitun, A. Bao, M. & Wang, K. L. Magnonic logic circuits. *J. Phys. D Appl. Phys.* **43**, 264005 (2010).

12 Manipatruni, S., Nikonov, D. E. & Young, I. A. Beyond CMOS computing with spin and polarization. *Nat. Phys.* **14**, 338-343 (2018).

13 Zografos, O. *et al.* Design and benchmarking of hybrid CMOS-spin wave device circuits compared to 10nm CMOS. *Proc. of the 15th IEEE Int. Conf. on Nanotechn.*, 686-689 (2015).

14 Duan Z. *et al.* Nanowire spin torque oscillator driven by spin orbit torques, *Nat. Commun.* **5**, 5616 (2014).

15 Wang, Q. *et al.* Spin pinning and spin-wave dispersion in nanoscopic ferromagnetic waveguides, *Phys. Rev. Lett.* **122**, 247202 (2019).

16 Wintz, S. *et al.* Magnetic vortex cores as tuneable spin-wave emitters, *Nat. Nano.*, **11**, 948-953 (2016).

17 Liu, C. *et al.* Current-controlled propagation of spin waves in antiparallel, coupled domains. *Nat. Nano.* **14**, 691-697 (2019).

18 P. Che. *et al.* Efficient wavelength conversion of exchange magnons below 100 nm by magnetic coplanar waveguides, *Nat. Commun.* **11**, 1445 (2020).

19 Wang, Q. *et al.* Reconfigurable nanoscale spin-wave directional coupler. *Sci. Adv.* **4**, e1701517 (2018).

20 Kirilyuk, A., Kimel, A.V. & Rasing, T. Ultrafast optical manipulation of magnetic order, *Rev. Mod. Phys.* **82**, 2731 (2010).

21 Kampfrath, T. *et al.* Coherent terahertz control of antiferromagnetic spin waves. *Nature Photon.* **5**, 31-34 (2011).

22 Lee, K. -S. & Kim, S, -K. Conceptual design of spin wave logic gates based on a Mach-Zehnder-type spin wave interferometer for universal logic functions. *J. Appl. Phys.* **104**, 053903 (2008).

23 Schneider, T., Serga, A. A. & Hillebrands, B. Realization of spin-wave logic gate. *Appl. Phys. Lett.*, **92**, 022505 (2008).

24 T. Goto, *et al.* Three port logic gate using forward volume spin wave interference in a thin yttrium iron garnet film. *Sci. Rep.* **9**, 16472 (2019).

25 Khivintsev, Y. V. *et al.* Spin waves in YIG-based networks: Logic and signal processing. *Phys. Metals Metallogr.* **120**, 1318 (2019).

26 Fischer, T. *et al.* Experimental prototype of a spin-wave majority gate. *Appl. Phys. Lett.* **110**, 152401 (2017).

27 Talmelli, G. *et al.* Reconfigurable nanoscale spin wave majority gate with frequency-division multiplexing, arXiv:1908.02546 (2019).

28 Wu, H. *et al.* Magnon valve effect between two magnetic insulators. *Phys. Rev. Lett.* **120**, 097205 (2018).

29 Vogt, K. *et al.* Realization of a spin-wave multiplexer, *Nat. Commun.* **5**, 3727 (2014).

30 Heussner, F. *et al.* Experimental realization of a passive gigahertz frequency-division demultiplexer for magnonic logic networks, *Phys. Stat. Sol.* **14**, 1900695 (2020).

31 Papp, A. *et al.* Nanoscale spectrum analyzer based on spin-wave interference, *Sci. Rep.* **7**, 9245 (2017).

32 Torrejon, J. *et al.* Neuromorphic computing with nanoscale spintronic oscillators. *Nature* **547**, 428-431 (2017).

33 Brächer, T. & Pirro, P. An analog magnon adder for all-magnonic neurons. *J. Appl. Phys.* **124**, 152119 (2018).

34 Chumak, A. V. Fundamentals of magnon-based computing, Spintronics Handbook: *Spin Transport and Magnetism. edited by E. Y. Tsymbal and I. Žutić*, p247-303, (2019).

35 Sadovnikov, A. V. *et al.* Directional multimode coupler for planar magnonics: Side-coupled magnetic stripes. *Appl. Phys. Lett.* **107**, 202405 (2015).

36 Sebastian, T. *et al.* Micro-focused Brillouin light scattering: imaging spin waves at the nanoscale. *Front. Phys.* **3**, 35 (2015).

37 Heinz, B. *et al.* Propagation of spin-wave packets in individual nanosized yttrium iron garnet magnonic conduits. *Nano Lett*. **20**, 4220-4227 (2020).

38 Clausen, P. et al, Mode conversion by symmetry breaking of propagating spin waves, *Appl. Phys. Lett.* **99**, 162505 (2011).

39 Brächer, T., Pirro, P. & Hillebrands, B. Parallel pumping for magnon spintronics: Amplification and manipulation of magnon spin currents on the micron-scale. *Phys. Rep.* **699**, 1-34 (2017).

40 Verba, R. *et al.* Amplification and stabilization of large-amplitude propagating spin waves by parametric pumping. *App. Phys. Lett.* **112**, 042402 (2018).

41 Mohseni, M. *et al.* Parametric generation of propagating spin waves in ultrathin yttrium iron garnet waveguides. *Phys. Status Solidi PRL* **14**, 2000011 (2020).

42 Imre, A. *et al.* Majority logic gate for magnetic quantum-dot cellular automata. *Science* **311**, 205-208 (2006).

43 Verba, R. *et al.* Excitation of propagating spin waves in ferromagnetic nanowires by microwave voltage-controlled magnetic anisotropy. *Sci. Rep.* **6**, 25018 (2016).

44 Schneider, M. *et al.* Bose-Einstein condensation of quasi-particles by rapid cooling. *Nat. Nano.* **15**, 457-461 (2020).

45 Bauer, H. G. *et al.* Nonlinear spin-wave excitations at low magnetic bias fields. *Nat. Comm.* **6**, 8274 (2015).

46 O. V. Dobrovolskiy, *et al.*, Spin-wave phase inverter upon a single nanodefect, *ACS Appl. Mater. Interf.* **11**, 17654 (2019).

47 Chen, Y. *et al.* Parametric resonance of magnetization excited by electric field. *Nano. Lett.* **17**, 572-577 (2017).

48 Sadovnikov, A. V. *et al.* Magnon straintronics: reconfigurable spin-wave routing in strain-controlled bilateral magnetic stripes. *Phys. Rev. Lett.* **120**, 257203 (2018).

49 Guillemard, C. et al. Ultralow magnetic damping in $Co_2Mn$-based Heusler compounds: promising materials for spintronics. *Phys. Rev. Appl.* **11**, 064009 (2019).

50 Robertson J., Liquid phase epitaxy of garnets, *J. Cryst. Growth* **45**, 233 (1978).

51 Maksymov, I. S. & Kostylev, M. Broadband stripline ferromagnetic resonance spectroscopy of ferromagnetic films, multilayers and nanostructures. *Physica E* **69**, 253-293 (2015).

52 Morozova M. A. *et al.* Suppression of periodic spatial power transfer in a layered structure based on ferromagnetic films, *J. Magn. Magn. Mater.* **466**, 119-124 (2018).

53 Vansteenkiste, A. *et al.* The design and verification of MuMax3. *AIP Adv.* **4**, 107133 (2014).

54 Kumar, D. *et al.* Numerical calculation of spin wave dispersions in magnetic nanostructures. *J. Phys. D: Appl. Phys.* **45**, 015001 (2012).

55 Venkat, G. Fangohr, H. & Prabhakar, A. Absorbing boundary layers for spin wave micromagnetics. *J. Magn. Magn. Mater.* **450**, 34-39 (2018).

**Acknowledgements:**

The authors thank Burkard Hillebrands for support and valuable discussions. This research has been supported by ERC Starting Grant 678309 MagnonCircuits, FET-OPEN project CHIRON (contract number 801055), by the Deutsche Forschungsgemeinschaft (DFG, German Research Foundation) – TRR 173 – 268565370 (Collaborative Research Center SFB/TRR-173 "Spin+X", Project B01) and through the Project 271741898, and by the Austrian Science Fund (FWF) through the project I 4696-N, and by the Ministry of Education and Science of Ukraine, Project 0118U004007. B. H. acknowledges support by the Graduate School Material Science in Mainz (MAINZ).

**Author contributions**

Q. W. proposed the directional coupler and half-adder design, preformed the BLS measurements, carried out the evaluation, and wrote the first version of the manuscript. C. D. provided the YIG film. M. K., B. H. and B. L. fabricated the nanoscale directional coupler. M. S., B. H. and M. G developed the BLS setup. T. B. acquired the SEM micrograph. M. K. performed the VNA-FMR measurements. R. V. developed the analytical theory and performed the theoretical calculations. Q. W. and M. M. performed the micromagnetic simulations. F. C., C. A. and S. D. C. performed the benchmarking and calculated the parameters of 7 nm CMOS half-adder. O. D. and T. B. discussed the interpretation and the relevance of the results. P. P. and A. V. C. led this project. All authors contributed to the scientific discussion and commented on the manuscript.

**Competing interests**

The authors declare no competing interests

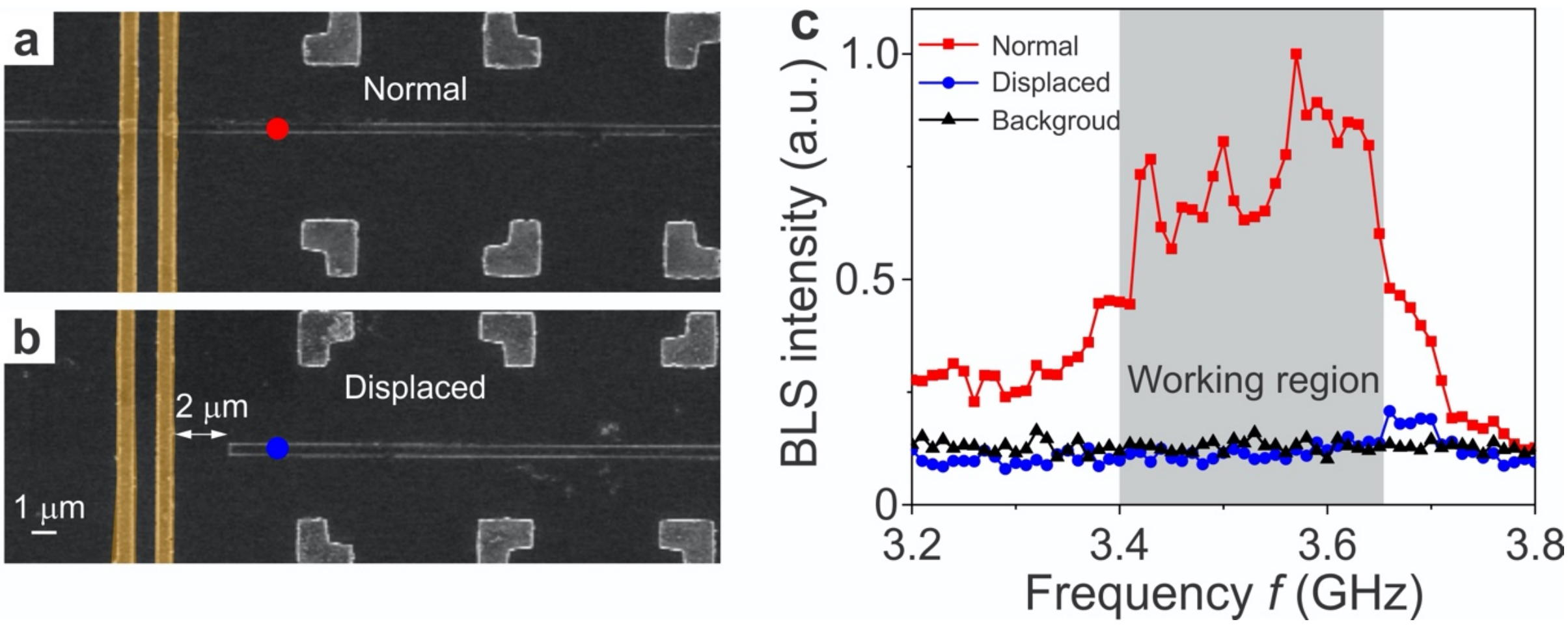

**Extended Data Fig. 1 Effect of far-field excitation by the U-shaped antenna.** *SEM images of the (**a**) normal and (**b**) displaced waveguides. The red and blue dots show the μBLS measurement points. **c**. The spin-wave intensities for normal (red dot line), displaced waveguides (blue dot line) and thermal background (black dot line). The grey area shows the working frequency range in the paper.*

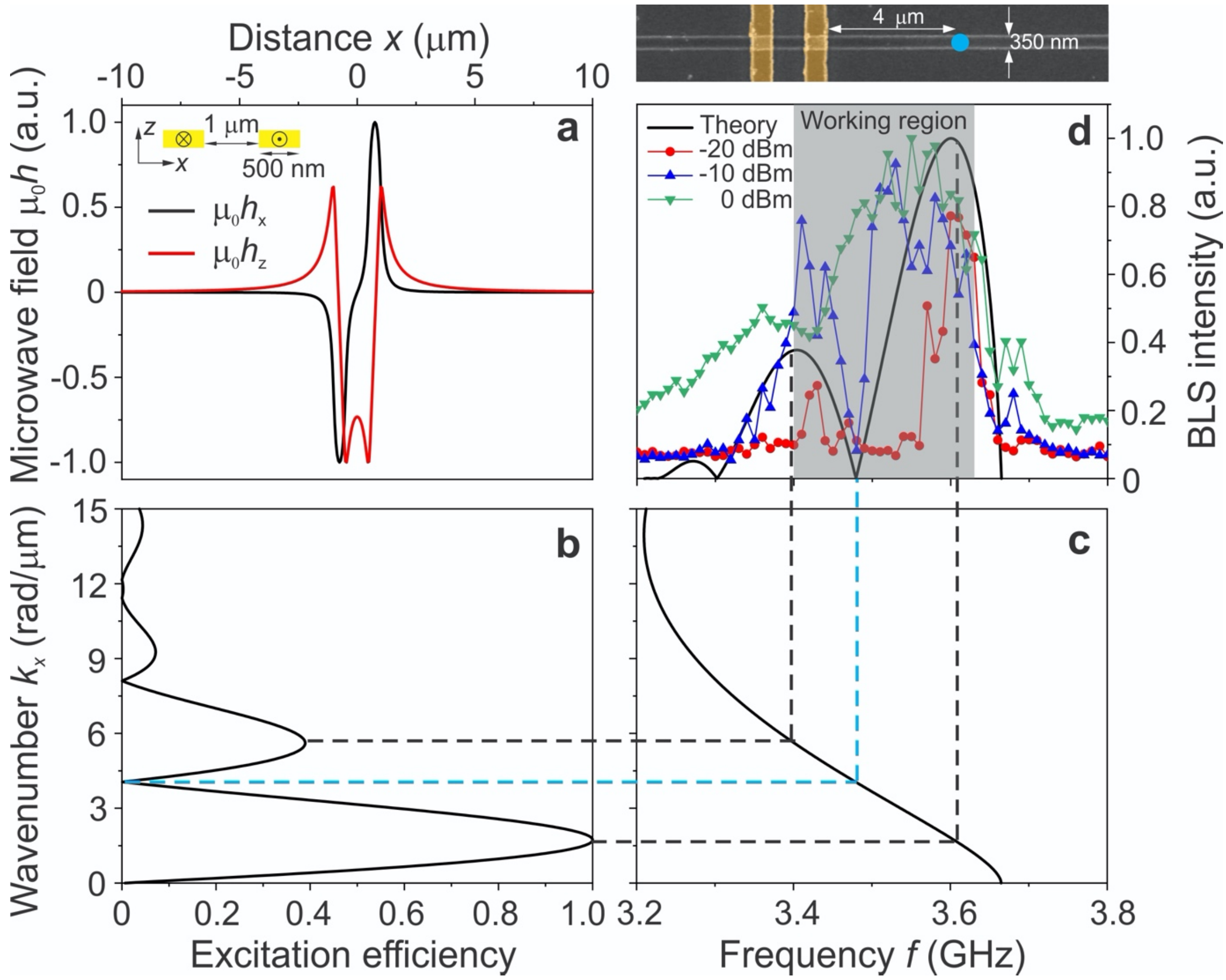

**Extended Data Fig. 2 Spin-wave spectra in the isolated waveguide.** ***a***. *The in-plane (black line) and out-of-plane (red line) field distribution created by the U-shape antenna. The schematic cross section of a U-shaped antenna is shown inset.* ***b***. *The excitation efficiency as a function of spin-wave wavenumber.* ***c***. *Spin-wave frequency as a function of spin-wave wavenumber.* ***d***. *Spin-wave intensities are measured 4 μm far from the antenna for different excitation powers. The black line shows the analytical calculation of the spin-wave intensity. A SEM image of the isolated waveguide is shown on the top of Fig. 2d.*

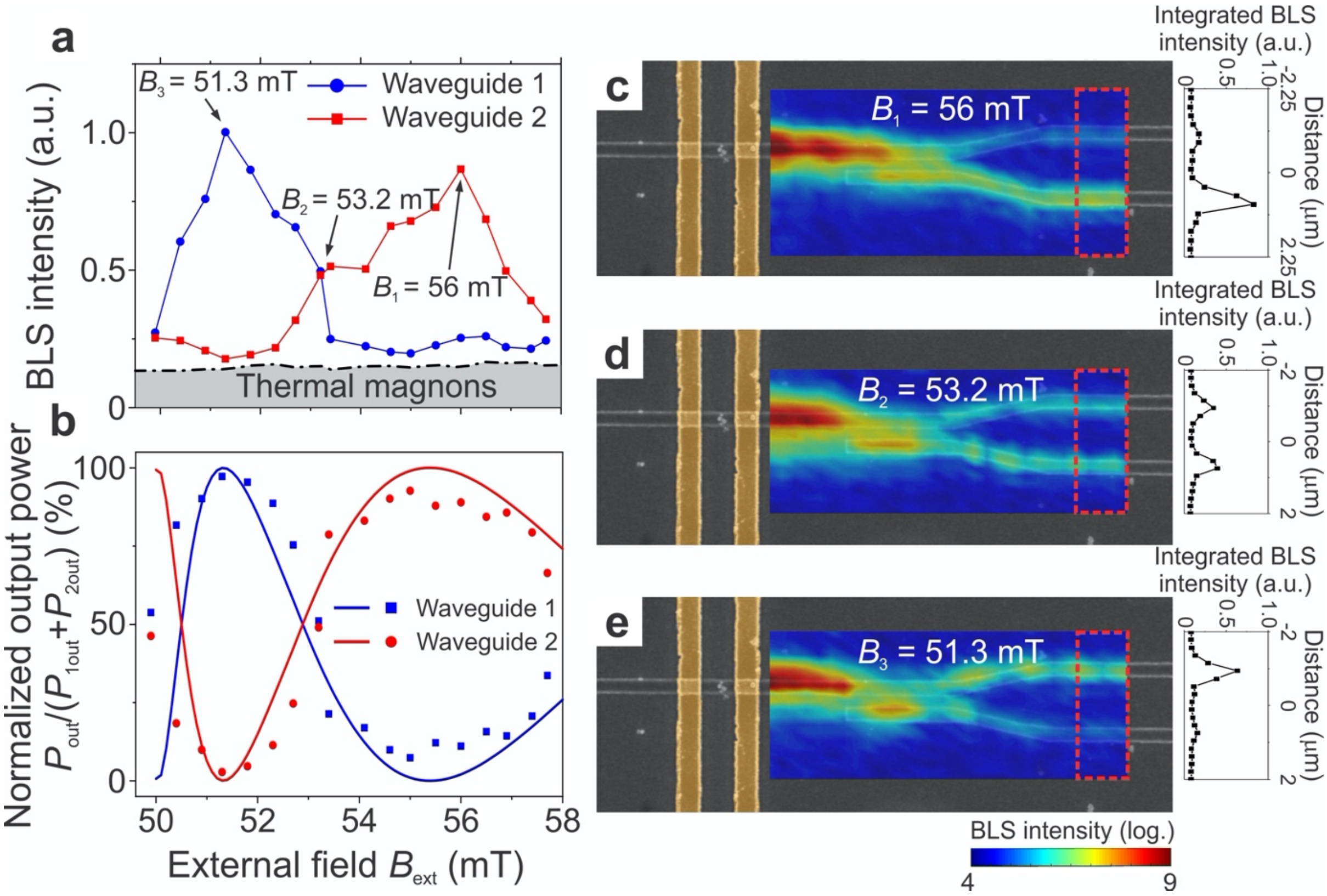

**Extended Data Fig. 3. Reconfigurability of the directional coupler by an applied magnetic field.** ***a***. *The averaged spin-wave intensity for a frequency of 3.465 GHz as function of the external field for the first (blue circles) and the second (red squares) output waveguide of the directional coupler.* ***b***. *Measured (circles and squares) and theoretically calculated (solid lines) normalized output spin-wave intensities at the first (blue) and second (red) output waveguide for different external fields.* ***c-e***: *Two-dimensional BLS maps of the spin-wave intensity for the different external magnetic fields:* **c.** *$B_1$ = 56 mT,* **d.** *$B_2$ = 53.2 mT, and* **e.** *$B_3$ = 51.3 mT. The right panels show the BLS intensity integrated over the red dashed rectangular regions.*

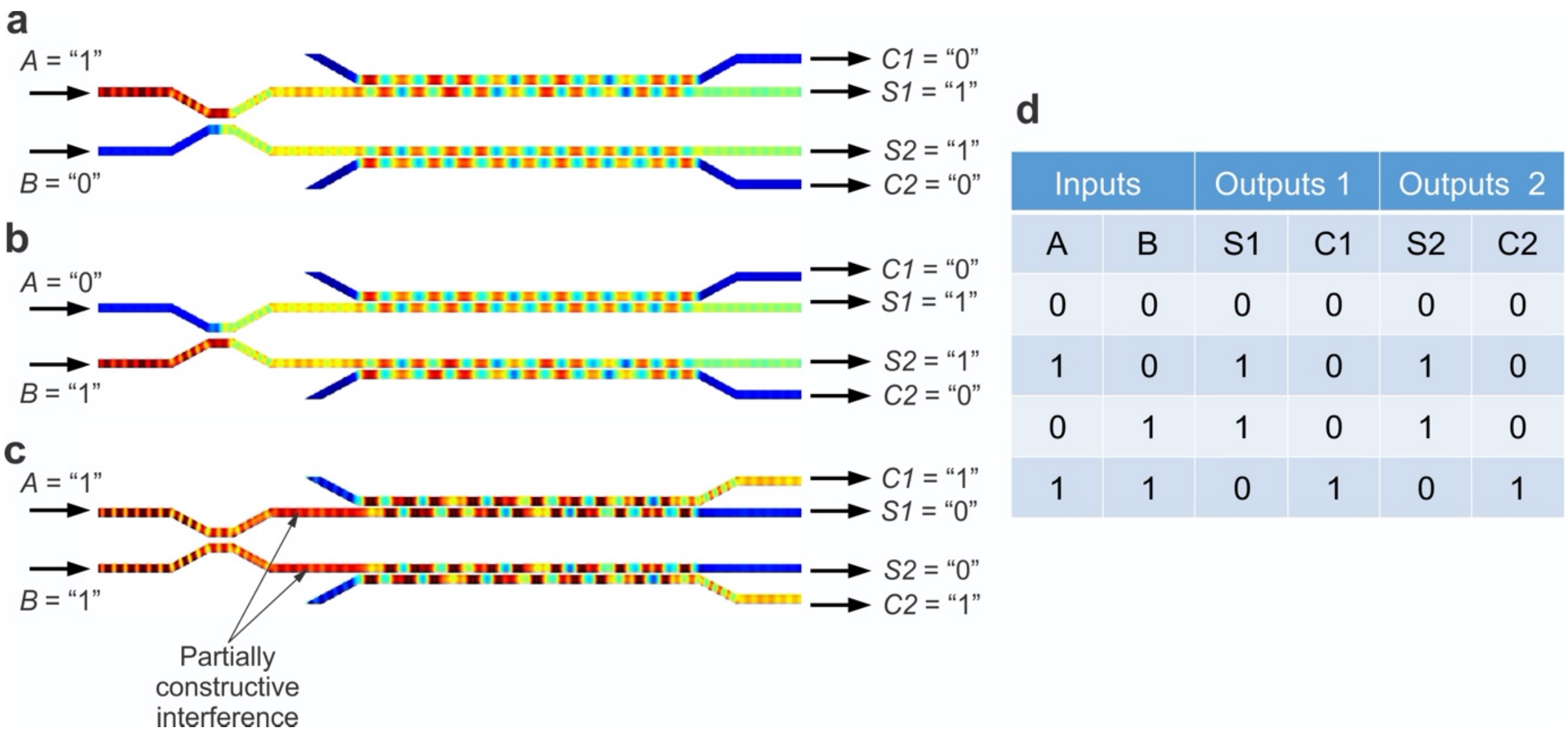

| Inputs | | Outputs 1 | | Outputs 2 | |
|---|---|---|---|---|---|
| A | B | S1 | C1 | S2 | C2 |
| 0 | 0 | 0 | 0 | 0 | 0 |
| 1 | 0 | 1 | 0 | 1 | 0 |
| 0 | 1 | 1 | 0 | 1 | 0 |
| 1 | 1 | 0 | 1 | 0 | 1 |

**Extended Data Fig.4. Modified half-adder with fan-out gate.** ***a-c*** *Operational principle and (d) truth table of two half-adders with shared inputs (corresponds to the half-adder with added fan-out logic gate).*

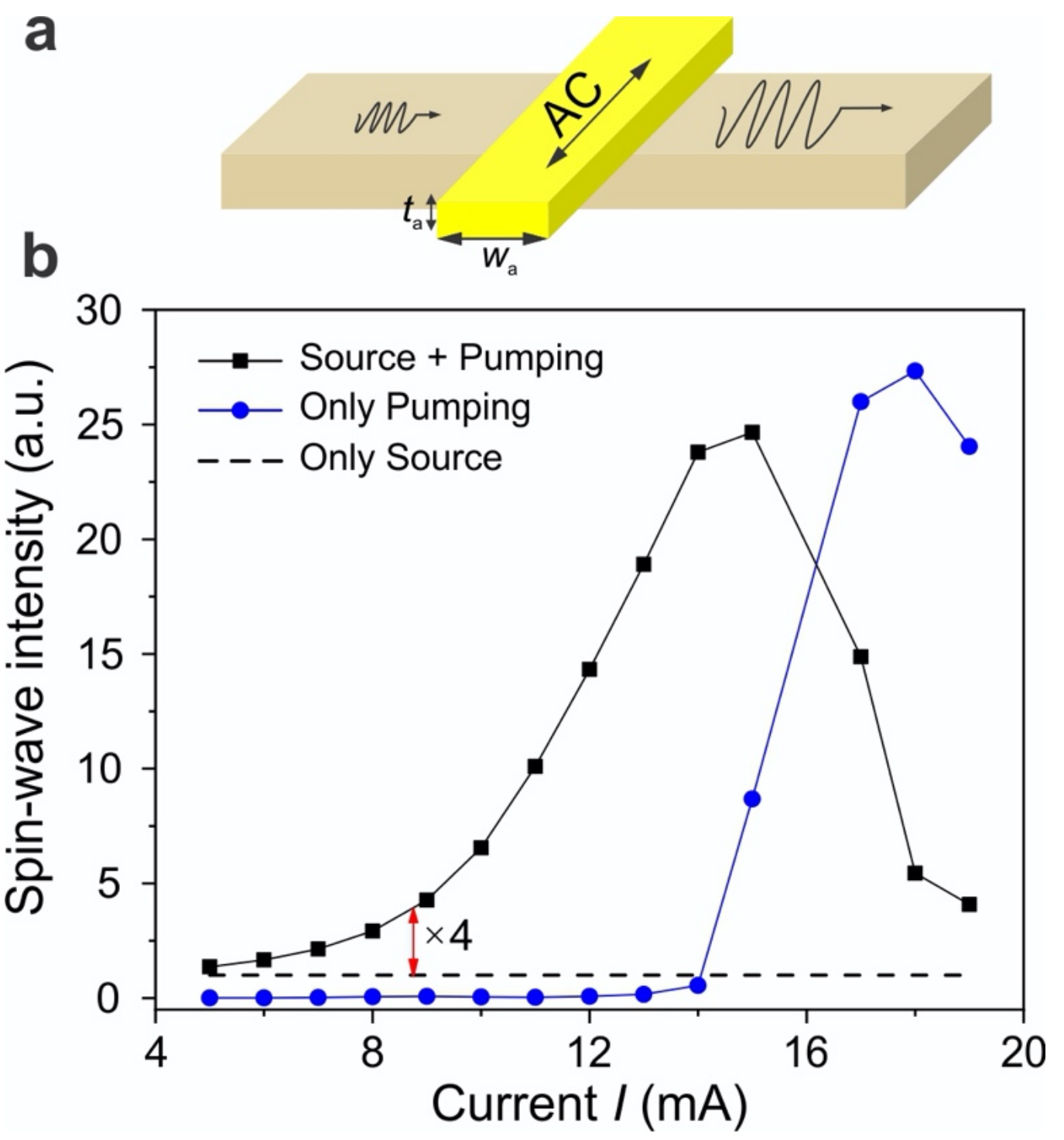

**Extended Data Fig. 5 Parametric amplification.** ***a*** *A schematic picture of the parametric amplifier.* ***b*** *Output spin-wave intensity as a function of pumping current for different conditions.*